\documentclass[aps,prl,showpacs,twocolumn]{revtex4-1}

\usepackage{amsmath}
\usepackage{graphicx}

\begin{document}

\title{Phase Transitions in Brownian Pumps}

\author{Marcel Dierl$^{1}$} 
\author{Wolfgang Dieterich$^{2}$} 
\author{Mario Einax$^{1,3}$} 
\author{Philipp Maass$^{1}$} 

\affiliation{
$^{1}$Fachbereich Physik, Universit\"at Osnabr\"uck,
  Barbarastra\ss e 7, 49076 Osnabr\"uck, Germany\\
$^{2}$Fachbereich Physik, Universit\"at Konstanz, 78457 Konstanz, Germany\\
$^{3}$School of Chemistry, Tel Aviv University, Tel Aviv 69978, Israel
}

\date{8 January 2013}

\begin{abstract}
  We study stochastic particle transport between two reservoirs along
  a channel, where the particles are pumped against a bias by a
  traveling wave potential. It is shown that phase transitions of
  period-averaged densities or currents occur inside the channel when
  exclusion interactions between the particles are taken into
  account. These transitions reflect those known for the asymmetric
  simple exclusion process (ASEP). We argue that their occurrence is a
  generic feature of Brownian motors operating in open systems.
\end{abstract}

%05.70.Ln Nonequilibrium and irreversible thermodynamics 
%05.40.-a Fluctuation phenomena, random processes, noise, and Brownian motion
%05.60.-k Transport processes

\maketitle

In connection with directed transport on the molecular level, two
research areas have attracted great attention in the past: Brownian
motors \cite{Juelicher/etal:1997, Astumian:1997, Reimann:2002,
  Astumian/Haenggi:2002, Kay/etal:2007, Haenggi/Marchesoni:2009}, and
driven diffusion systems under a static bias \cite{Derrida:1998,
  Schuetz:2001, Golinelli/Mallick:2006,
  Blythe/Evans:2007,Kolomeisky:2013}. Brownian motors are operated by
a periodic process in time, where, in contrast to classical engines,
fluctuations caused by thermal noise and thermally assisted overcoming
of energy barriers are important \cite{Magnasco:1993,
  Astumian/Haenggi:2002}. They exhibit some sort of symmetry breaking
as a genuine feature \cite{Haenggi/Marchesoni:2009}, which is
essential for their functioning. Driven diffusion under a static bias
has received particular interest in connection with transport through
open tube-like compartments because of boundary-induced phase
transitions \cite{Krug:1991, Klumpp/Lipowsky:2003}. Essential for
their occurrence is the consideration of exclusion interactions, which
means that two particles cannot occupy the same place.

We show in this work that boundary-induced phase transitions also
appear in cyclically operating Brownian pumps, if exclusion
interactions are taken into account. As argued further below, the
occurrence of boundary-induced phase transitions should be generic for
Brownian motors with interacting particles in open environments. For
demonstration we consider a model for a traveling wave potential
\cite{note:motordef}, first introduced by Dhar and coworkers
\cite{Chaudhuri/Dhar:2011, Jain/etal:2007}.

\begin{figure}[b!]
\includegraphics[width=0.43\textwidth]{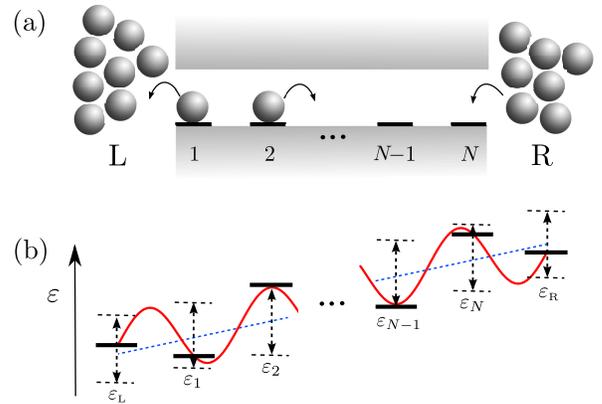}
\caption{\label{fig:model} (color online). (a) Sketch of a
   channel with $N$ equidistant sites in contact with two particle
   reservoirs L and R to the left and right. (b) Illustration of the
   site energies $\varepsilon_i$ along the channel. The dashed blue
   line marks a linear increase of the (time-averaged) site energies
   from left to right by a bias $F>0$.  A time-dependent traveling
   wave potential allows for pumping of particles against the
   bias. Its superposition with the bias potential yields the total
   potential marked by the red solid line.}
\end{figure}

The model is sketched in Fig.~\ref{fig:model}. A channel with $N$
sites connects two reservoirs L and R to the left and right with
particle densities $\rho_L$ and $\rho_R$. Particles from these
reservoirs are both injected into and ejected from the channel. The
sites $i=1,\ldots, N$ can be occupied by at most one particle and a
particle occupying site $i$ has energy $\varepsilon_i$. Likewise, a
particle in the reservoirs L or R has energies
$\varepsilon_L=\varepsilon_0$ or $\varepsilon_R=\varepsilon_{N+1}$.
The particles inside the channel perform thermally activated jumps to
vacant nearest-neighbor sites with rates
$\Gamma_{i,i\pm1}=\nu\exp[-(\varepsilon_{i\pm1}-\varepsilon_i)/2k_BT]$,
where $\nu$ is an attempt frequency and $k_BT$ the thermal energy.  In
the following, we choose $\nu^{-1}$ and $k_BT$ as our time and energy
unit ($\nu=1$, $k_BT=1$) and the site-to-site distance $a$ as length
unit ($a=1$). Along the channel, a constant bias $F$ is present,
leading to site energies linearly increasing from left to
right. Accordingly, for $F>0$, a net particle flow from R to L occurs.

%------ FIGURE 1 --------
%------------------------

Application of a traveling wave potential from L to R can generate a
particle flow against the bias. The potential modulates the site
energies periodically in time and space with period $\tau$ and
wavelength $\lambda$, yielding
\begin{equation}
\varepsilon_i(t)=iF+
A\sin\left(\frac{2\pi i}{\lambda}-\frac{2\pi t}{\tau}\right)\,,\quad
i=0,\ldots,N+1\,,
\label{eq:epsi}
\end{equation}
where $A$ is the modulation amplitude. Because of the cyclic
modulation of the reservoir energies and the site energies next to the
reservoirs, the injection and ejection rates, irrespective of their
detailed functional form, also vary periodically in time.

In the stationary state, the period-averaged net current $\bar J$
between two neighboring sites must be the same along the channel
because of particle number conservation. Period-averaged particle
densities $\bar\rho_i$ at sites $i$ are not spatially homogeneous. The
form of the density profile depends on details of system-reservoir
couplings specified by the injection and ejection rates. For the
period-averaged densities and currents we hence encounter a situation
analogous to driven diffusion in an open system under static bias:
Phase transitions of the bulk density $\bar\rho_B$ inside the channel
should occur upon changing the densities $\rho_L$ and $\rho_R$ in the
left and right reservoir.

\begin{figure}[t!]
\includegraphics[width=0.41\textwidth]{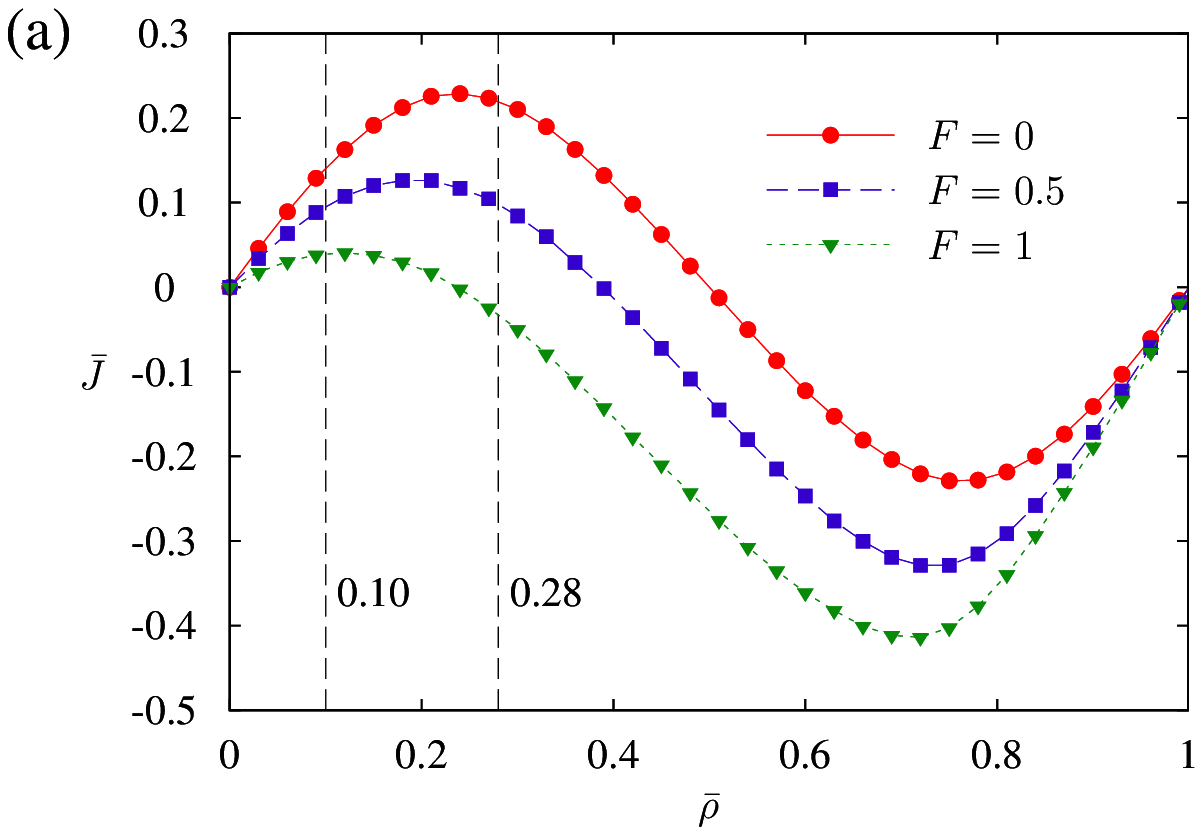}\\[1ex]
\includegraphics[width=0.41\textwidth]{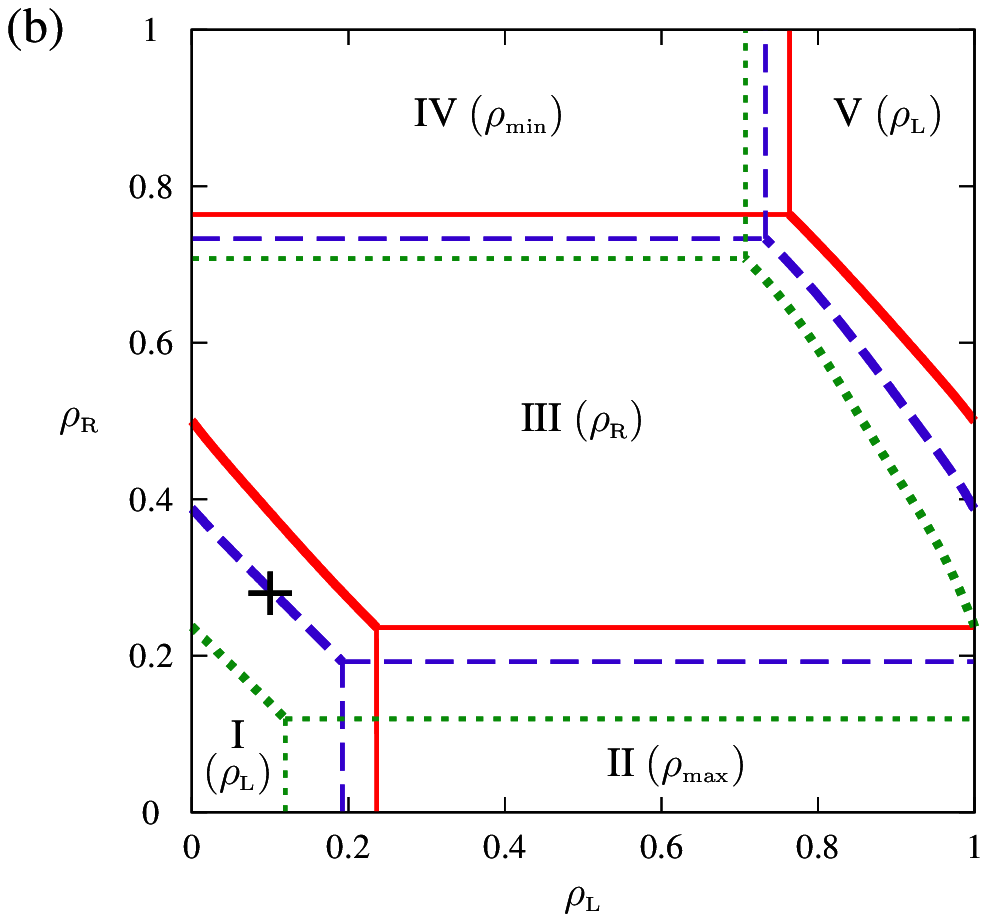}
\caption{\label{fig:j-rho} (color online). (a) Period-averaged current
  $\bar J$ as a function of period-averaged density $\bar\rho$ in a
  bulk system for three different $F$ values, and fixed parameters
  $\lambda=4$, $\tau=2$, and $A=3$ of the traveling wave
  potential. Data were obtained by KMC simulations of a closed
  channel. (b) Phase diagram derived from (a) by applying the
  minimum/maximum current principles. Phase transitions of first order
  are marked by thick lines, and transitions of second order by thin
  lines. The assignment of line types (and colors) to the $F$ values
  is the same as in the legend of (a). The five different phases are
  labeled by roman numbers with respective bulk densities given in
  brackets. The cross marks the point $(\rho_L,\rho_R)=(0.1,0.28)$
  used in the example for demonstrating the influence of a phase
  transition on $\bar J$ in Fig.~\ref{fig:j-f}; in (a) the densities
  $0.1$ and $0.28$ are marked by vertical dashed lines.}
\end{figure}

An overview of the occurring phases is obtained based on the bulk
current-density relation.  This relation applies to channels with
periodic boundary conditions, where the density $\bar\rho$ is
homogeneous. In the following we refer to those as ``closed
channels''.  Figure~\ref{fig:j-rho}(a) shows the bulk current-density
relation for three different values of $F=0$, 1/2, and 1, and some
fixed values of the other parameters.  The results were obtained by
kinetic Monte Carlo (KMC) simulations for time-dependent rates as
described in \cite{Holubec/etal:2011}. For all $F$, there exists a
density $\rho_0(F)$, where $\bar J$ changes sign. Current reversals
have often been observed in Brownian motors, both in many-particle
\cite{Derenyi/Vicsek:1995, Chaudhuri/Dhar:2011, Jain/etal:2007} and in
one-particle models \cite{Reimann/Haenggi:2002, Reimann:2002,
  Kostur/Luczka:2001}. For $\bar\rho<\rho_0$, $\bar J$ goes through a
maximum at $\bar\rho=\rho_{\rm max}$, and for $\bar\rho>\rho_0$
through a minimum at $\bar\rho=\rho_{\rm min}$. The densities
$\rho_0$, $\rho_{\rm max}$, and $\rho_{\rm min}$ all decrease with
increasing $F$. In the absence of the bias, $F=0$, a sinusoidal
behavior of $\bar J(\bar \rho)$ is found, and $\bar J(\bar\rho)=-\bar
J(1-\bar\rho)$.  This property is a consequence of particle-hole
symmetry, and the sinusoidal form can be understood from a
perturbative treatment in the wave amplitude $A$
\cite{Chaudhuri/Dhar:2011, Jain/etal:2007}.

%------ FIGURE 2 --------
%------------------------

\begin{figure}[t!]
\includegraphics[width=0.41\textwidth]{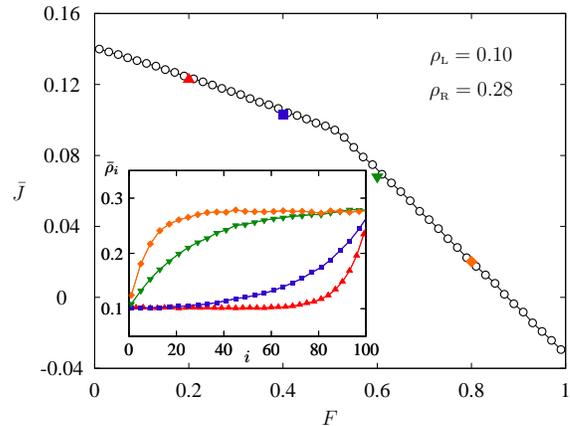}
\caption{\label{fig:j-f} (color online). Current $\bar J$ as
  function of $F$ for fixed reservoir densities and the same
  parameters of the traveling wave potential as in
  Fig.~\ref{fig:j-rho}.  The open circles are for closed channels. The
  filled colored symbols refer to $\bar J$ in open channels at
  $F=0.2$, 0.4, 0.6, and 0.8, for which the respective density
  profiles are shown in the inset.}
\end{figure}

Let us now consider the open channel in Fig.~\ref{fig:model}.  In this
open channel the particle number is fluctuating in the stationary
state and density profiles $\bar\rho_i$ form with a flat regime
$\bar\rho_i=\bar\rho_B$ inside the channel (see inset of
Fig.~\ref{fig:j-f} for examples).  The density $\bar\rho_B$ and,
accordingly, the appearance of the different phases depends on the
type of system-reservoir couplings \cite{Dierl/etal:2012,
  Dierl/etal:2013}. Their identification in general requires the
scanning of all possible values of $\rho_L$ and $\rho_R$ for the open
channel. However, for a particular choice of coupling, coined
``bulk-adapted'', $\bar\rho_B$ can be derived from the curves in
Fig.~\ref{fig:j-rho}(a) by applying minimum/maximum current
principles. These were first introduced for diffusion systems under
static bias \cite{Krug:1991, Popkov/Schuetz:1999, Hager/etal:2001} and
it is rather surprising at first sight that they can be taken over to
time-varying driving forces. The reason is that, as under static bias,
a continuity equation relates the period-averaged densities and
currents, with a source-free $\bar J$ in the stationary
regime. Applied to the period-averaged quantities, the principles
state that if $\rho_L$ is smaller (larger) than $\rho_R$, the bulk
density $\bar\rho_B$ equals the density, where $\bar J(\bar\rho)$
assumes its minimal (maximal) value in the intervals
$\rho_L<\bar\rho<\rho_R$ ($\rho_R<\bar\rho<\rho_L$). This is a
consequence of the fact that, to match the reservoir densities at the
boundaries, the period-averaged density profiles cannot be uniform,
and accordingly diffusive currents occur. These diffusive currents
plus $\bar J(\bar\rho)$ must be constant everywhere. For example, if,
for $\rho_L<\rho_R$, $\bar\rho_i$ increases monotonically from left to
right, the diffusive current is negative, or zero for
$\bar\rho_i=\bar\rho_B$. Accordingly, to keep the sum of $\bar
J(\bar\rho)$ and the diffusive current constant, $\bar J(\bar\rho)$
must be minimal in the region of flat density $\bar\rho_B$. The
technical details for implementing the bulk-adapted couplings are
described in \cite{suppl:ba-couplings}.

Applying the minimum/maximum current principles to the curves in
Fig.~\ref{fig:j-rho}(a) yields the five phases shown in
Fig.~\ref{fig:j-rho}(b), where $\bar\rho_B$ equals either $\rho_L$
(phases I, V), $\rho_R$ (phase III), $\rho_{\rm max}$ (phase II) or
$\rho_{\rm min}$ (phase IV). First-order transitions between two
phases with jumps of $\bar\rho_B$ are marked by thick lines, and
second-order transitions are marked by thin lines. These transition
lines shift when $F$ changes.  In phases I and V the density profiles
are flat on the left side (cp.\ the profiles for $F=0.2$ and 0.4 in
the inset of Fig.~\ref{fig:j-f}), in phase III on the right side (cp.\
the profiles for $F=0.6$ and 0.8 in the inset of Fig.~\ref{fig:j-f}),
and in the maximum and minimum current phases II and IV, the density
profile is flat in the channel's interior and bent towards $\rho_L$
and $\rho_R$ at the two boundaries.

The phase transitions influence the current $\bar J$ and hence the
efficiency of the Brownian pump. To demonstrate this, let us
consider, as an example, the particle flow against increasing bias
$F$, for fixed reservoir densities $\rho_L=0.1$ and $\rho_R=0.28$, and
the same parameters of the traveling potential as in
Fig.~\ref{fig:j-rho}.  The corresponding point
$(\rho_L,\rho_R)=(0.1,0.28)$ is marked by the cross in
Fig.~\ref{fig:j-rho}(b). Starting from $F=0$, this point lies in phase I
($\bar\rho_B=0.1$), and at large $F=1$ it lies in phase III
($\bar\rho_B=0.28$). The transition occurs at $F=F_\star$ with
$F_\star\simeq0.5$. Density profiles from KMC simulations before
($F<F_\star$) and after this transition ($F>F_\star$) are shown in the
inset of Fig.~\ref{fig:j-f}. At $\bar\rho=0.1$, the current $\bar J$
in Fig.~\ref{fig:j-rho}(a) is more weakly varying with $F$ than at
$\bar\rho=0.28$. Accordingly, $\bar J$ as a function of $F$ displays a
kink at the transition point $F=F_\star$, see Fig.~\ref{fig:j-f}. The
open symbols in the figure are from KMC simulations of closed channels
with $\bar\rho=0.1$ for $F<F_\star$ and $\bar\rho=0.28$ for
$F>F_\star$. The four filled symbols are the currents in the open
channels with the density profiles shown in the inset. The agreement
of $\bar J$ for the open and closed channel confirms the validity of
the minimum/maximum current principles.

The effect of the shifting of phase transition lines on $\bar J$ has
been demonstrated here for a change of the bias. A shifting of
transition lines with corresponding effects on $\bar J$ can also be
induced by a variation of the parameters of the traveling
potential. This is relevant for the efficiency and optimization of the
pump. The occurrence of minimum and maximum current phases has an
interesting implication for the robustness of the motor's performance
against changes of the reservoir densities, because in these phases
$\bar\rho_B$ and hence $\bar J$ are independent of $\rho_L$ and $\rho_R$
\cite{note:j-2nd-order}.

For Brownian motors of other type, as, for example, a flashing or
rocking ratchet, analogous features are to be expected, if they
operate in an open environment. This is because, after
period-averaging of currents and densities, a net driving force (for
the motor to transport particles) and local particle number
conservation (leading to a coupling between currents and densities)
are present. Phase transitions will appear also for Brownian motors
treated in continuum, if particles with some hard-sphere diameter are
considered. In fact, the model in Fig.~\ref{fig:model} can be easily
connected to a discrete implementation of a Langevin equation
\cite{Chaudhuri/Dhar:2011}.  For particle interactions beyond site
exclusion, more complex phase diagrams can occur
\cite{Dierl/etal:2012, Hager/etal:2001}. Likewise, richer phase
diagrams can emerge by injection and ejection processes along
filaments according to a Langmuir kinetics
\cite{Parmeggiani/etal:2003}.

%------ FIGURE 3 --------
%------------------------

%%%%%%%%%%%%%%%%%%%%%%%%%%%%%%%%%%%%%%%%%%%%%%%%%%%%%%%%%%%%%%%%%%%

%\bibliography{lit}

\begin{thebibliography}{28}
\bibitem [{not({\natexlab{a}})}]{note:motordef}%
  \BibitemOpen
  \href@noop {} {}\bibinfo {note} {Sometimes the term
  `Brownian motors' is reserved for those motors, where thermal noise
  facilitates directed particle motion. The traveling wave potential is not of
  that sort, but the general concepts presented in this work equally apply to
  them.}\BibitemShut {Stop}%
\bibitem [{not({\natexlab{b}})}]{note:j-2nd-order}%
  \BibitemOpen
  \href@noop {} {}\bibinfo {note} {For transitions into and
  out of the minimum and maximum current phases induced by one (or several)
  varying control parameters, the currents before and after the transition have
  a different functional dependence on these parameters. The functional forms
  cross with continuous first-order derivative, but there will be singularities
  in higher-order derivatives.}\BibitemShut {Stop}%
\end{thebibliography}
%\bibliographystyle{apsrev4-1}


\end{document}